\documentclass{article} 

\usepackage[authoryear]{natbib}
\setcitestyle{authoryear}
\usepackage{authblk}

\usepackage[pageanchor=false]{hyperref}
\usepackage{lineno}
\usepackage{amsmath} 
\usepackage{amssymb}  
\usepackage{url}
\usepackage{subcaption}
\usepackage{mathtools}
\usepackage{bm}

\usepackage{rotating}
\usepackage{multirow}

\modulolinenumbers[5]

\newtheorem{remark}{Remark}{\normalfont}{\normalfont}

\newcommand{\So}{{\mathcal{S}}}            
\newcommand{\M}{{M}}             
\newcommand{\MS}{\mathcal{M}}    

\newcommand{\NN}{\mathcal{N}} 
\newcommand{\batchsize}{q}
\newcommand{\seqlen}{m}
\newcommand{\nin}{n_u} 
\newcommand{\ny}{n_y} 
\newcommand{\nx}{n_x}
\newcommand{\numiter}{n}
\newcommand{\nsamp}{N}
\newcommand{\tens}[1]{\bm{#1}}

\newcommand{\ntheta}{n_\theta}
\newcommand{\Yid}{Y}
\newcommand{\Uid}{U}
\newcommand{\Did}{{\mathcal{D}}}
\newcommand{\simul}[1]{{#1}^{\rm sim}}
\newcommand{\hidden}[1]{\overline{#1}}

\newcommand{\ymodel}{y^{\rm o}}

\newcommand{\norm}[1]{\left\lVert#1\right\rVert}
\newcommand{\est}{}

\usepackage{algorithm}

\usepackage{enumitem}
\renewcommand{\theenumi}{\arabic{enumi}}

\renewcommand{\theenumii}{\arabic{enumii}}

\usepackage{color}

\title{Continuous-time system identification with neural networks: model structures and fitting criteria} 
\author{Marco Forgione}
\author{Dario Piga}
\affil{IDSIA Dalle Molle Institute for Artificial Intelligence USI-SUPSI, Via la Santa 1, CH-6962 Lugano-Viganello, Switzerland.}

\begin{document}
\maketitle

\begin{abstract}
This paper presents tailor-made neural model structures and two custom fitting criteria for learning dynamical systems. The proposed framework is based
on a representation of the system behavior in terms of continuous-time state-space models. The sequence of hidden states is optimized along with the neural
network parameters in order to minimize the difference between measured and
estimated outputs, and at the same time to guarantee that the optimized state
sequence is consistent with the estimated system dynamics. The effectiveness
of the approach is demonstrated through three case studies, including two public system identification benchmarks based on experimental data.
\end{abstract}

\noindent\rule{\textwidth}{1pt}
 To cite this work, please use the following bibtex entry:
\begin{verbatim}
@article{forgione2021a,
  title={Continuous-time system identification
  with neural networks: Model structures and fitting criteria},
  author={Forgione, M. and Piga, D.},
  journal={European Journal of Control},
  volume={59},
  pages={69--81},
  year={2021},
  publisher={Elsevier}
}
\end{verbatim}
\vskip 1em
Using the plain bibtex style, the bibliographic entry should look like:\\ \\
\textsf{M. Forgione and D. Piga. Continuous-time system identification
  with neural networks: Model structures and fitting criteria.} \textit{European Journal of Control}, 59:69--81, 2021.

\noindent\rule{\textwidth}{1pt}

\section{Introduction}
In recent years, deep learning has advanced at a tremendous pace and is now the core methodology behind cutting-edge technologies such as speech recognition, image classification and captioning, language translation, and autonomous driving \citep{schmidhuber2015deep, lecun2015deep, bengio2009learning}. These impressive achievements are attracting ever increasing investments both from the private and the public sector, fueling further research in this field. 

A good deal of the advancement in the deep learning area is publicly accessible, both in terms of scientific publications and software tools.
For instance, highly optimized and user-friendly deep learning frameworks are readily available \citep{paszke:2017automatic, tensorflow2015-whitepaper-short}, and are often distributed under permissive open-source licenses.
Using the high-level functionalities of a deep learning framework and following good practice, even a novice user can tackle  \emph{standard}  machine learning tasks (once considered extremely hard) such as image classification with moderate effort. 

An experienced practitioner can employ the same deep learning framework at a lower level to tackle non-standard learning problems, by defining customized models and objective functions to be optimized, and using operators such as \emph{neural networks} as building blocks. 
The practitioner is free from the burden of writing optimization code from scratch for every particular problem, which would be tedious and error-prone.
In fact, as a built-in feature, modern deep learning engines can compute the derivatives of a supplied objective function with respect to free tunable parameters by implementing the celebrated \emph{back-propagation} algorithm \citep{rumelhart1988learning}. In turn, this enables convenient setup of any gradient-based optimization method.

An exciting, challenging---and yet largely unexplored---application field is \emph{system identification} with tailor-made model structures and fitting criteria. In this context, neural networks can be used to describe uncertain components of the  dynamics, while retaining structural (physical) knowledge, if available. Furthermore, the fitting criterion can be specialized to take into account the modeler's ultimate goal, which could be prediction, failure detection, state estimation, control design, simulation, \emph{etc}.  

The choice of the cost function may also be influenced by computational considerations. 
In this paper, in particular, models are assessed in terms of their simulation performance. In this setting, from a theoretical perspective, simulation error minimization is generally the best fitting criterion. However, evaluating the simulation error loss and its derivative may be prohibitively expensive from a computational perspective for dynamical models containing neural networks. Furthermore, simulation over time has an intrinsically sequential nature and offers limited opportunities for parallelization.

In this paper, we present two fitting algorithms 
whose run-time is significantly faster than full simulation error minimization, but still provide models with high simulation performance. 

In the first approach, called  \emph{truncated simulation error minimization}, the neural dynamical model is simulated over (mini)batches of 
\emph{subsequences} extracted from the training dataset. This allows parallelization of the model simulation over the different subsequences in a batch and also results in reduced back-propagation cost with respect to a full open-loop simulation. 
Special care is taken to provide consistent initial conditions for all the simulated subsequences. In fact, these initial conditions are optimized along with the neural network parameters according to a dual objective. Specifically, the batch cost function takes into account both the distance between the simulated output and the measured output---the fitting objective---and the consistency of all the initial condition variables with the neural model equation---the initial state consistency objective. This cost function is iteratively minimized over the randomly extracted batches through a gradient-based optimization algorithm.
Preliminary ideas of this approach are presented  in \cite{forgione20Uz} for discrete-time 
neural model structures and are extended here to the continuous-time case, with dynamics described in terms of Ordinary Differential Equations (ODE).

In the second approach, called \emph{soft-constrained integration},   
the neural dynamical model is enforced by a regularization term in the cost function penalizing the violation of a numerical ODE integration scheme applied to the system's (hidden) state variables.
These state variables, together with the neural network parameters, are tuned with the dual objective of fitting the measured data and minimizing the penalty term associated with the violation of the numerical integration scheme.
In the soft-constrained integration method, simulation through time is thus completely circumvented and the loss function splits up into independent contribution for each time step. This enables a fully parallel implementation of gradient-based optimization.

{\subsection{Related works}}
The use of neural networks in system identification has a long history. For instance, neural \emph{AutoRegressive Moving Average with eXogenous inputs} models are discussed  in \cite{chen1990non}. Training is performed with the one-step prediction error method \citep{ljung1978convergence} previously developed in the context of linear dynamical systems.
In \citep{horne1995experimental}, several \emph{Recurrent Neural Network} structures trained by \emph{Back-Propagation Through Time} are evaluated on a nonlinear system identification task. {Minimization of the simulation error loss is performed over short training sequences, with known initial conditions.}
Although the overall reasoning in these earlier works is similar to our, their results are hardly comparable, given the huge gap of hardware/software technology.

More recently, a few interesting approaches exploiting modern deep learning concepts and tools for system identification have been proposed. {For instance, \cite{masti2018learning} introduces a technique to identify neural state-space model structures using deep autoencoders for state reconstruction}, while \cite{gonzalez2018non} and \cite{wang2017new} discuss the use of \emph{Long Short-Term Memory} (LSTM) recurrent neural networks for system identification. {A recent comparative study on modern neural architectures for system identification is also presented in \cite{kumar2019comparative}}.  

Compared to these recent contributions, our work focuses on specialized neural model structures for the identification task at hand. 
Furthermore, all the above mentioned contributions consider discrete-time (DT) dynamical systems, while we deal with the continuous-time (CT) case in this work.

{
Nowadays, direct identification of CT dynamical systems is an active research topic attracting the attention of many researchers in the system and control community. 
There are indeed multiple advantages in direct CT identification, as argued in~\cite{Hugues2014}.  
First, the majority of physical systems are naturally modeled in a continuous-time framework. 
Embedding physical knowledge in continuous-time model structures is thus more intuitive, and inspecting a continuous-time identified model is more insightful as some parameters may retain a physical meaning.
Second, continuous-time models can handle the case of non-uniformly sampled data. Last, continuous-time identification is generally immune from the numerical issues affecting discrete-time methodologies in the case of high sampling frequency. 
Complete reviews of direct CT identification algorithms and applications  can be found in the papers~\cite{garnier2003continuous,mercere2011continuous,garnier2015direct,Hugues2014,LaPiPiTo17,PigaCont}, in   the book~\cite{Garnier:2008:ICM:1796449}, and in the special issue~\cite{garnier2014special}. However, the above contributions only focus on   dynamical systems with linear relationships in the  input and output signals, such as linear time-invariant, linear parameter-varying and linear time-varying systems. Identification of general nonlinear CT dynamical  models is still a challenging and open    problem.   
}

{A few contributions have tackled non-linear CT system identification using certain special neural model structures. For instance, in  \cite{liu2008continuous} an identification scheme for CT dynamical systems described in a non-linear observability canonical form is proposed, while \cite{ren2003identification} consider dynamic neural networks based on feedback linearization theory. Both methods require high-order derivatives of the system output for training, whose estimate could be problematic in the presence of measurement noise. 
}


The connection between deep learning and dynamical system theory is currently under intensive investigation, see \emph{e.g.}, \citep{weinan2017proposal} and cross-contamination is yielding substantial advances to both fields. 
On the one hand, certain modern deep learning architectures are now interpreted as discrete-time approximations of an underlying continuous-time neural dynamical system. {Equivalently, continuous-time neural network architectures like the ones considered for modeling in this paper are seen as the limit case of deep learning architectures for an infinite number of layers. This is discussed, for instance, in the introduction of \cite{chen2018neural}.}
Exploiting this parallel, \citep{haber2017stable} and \citep{ruthotto2019deep} analyze the stability properties of existing deep neural architectures through the lens of system theoretic tools. Modified  architectures guaranteeing stability \emph{by design} are also proposed.
On the other hand, contributions such as \cite{raissi2019physics, long2019pde, rackauckas2020universal} showcase the potential of neural networks for data-driven modeling of dynamical systems described by ordinary and partial differential equations. With respect to these contributions, our aim is to devise computationally efficient fitting strategies for neural dynamical models that are robust to the measurement noise.

{\subsection{Paper structure}}
The  rest of this paper is structured as follows. The overall settings and problem statement is outlined in Section \ref{sec:settings}. The neural dynamical model structures are introduced in Section \ref{sec:model_structure} and   criteria for fitting these model structures to training data are described in Section \ref{sec:training}. Simulation results are presented in Section \ref{sec:example} and can be replicated using the codes available at \url{https://github.com/forgi86/sysid-neural-continuous}.  Conclusions and directions  for future research are  discussed in Section \ref{sec:conclusions}. 

\section{Problem Setting}
\label{sec:settings}
We are given a {training} dataset $\Did$ consisting of $\nsamp$ {sequential} input samples $\Uid = \{u_{0},\;u_{1},\dots,\;u_{\nsamp-1}\}$ and output samples  $\Yid = \{y_{0},\;y_{1},\dots,\;y_{\nsamp-1}\}$, gathered at time instants $T = \{t_0=0,\; t_1,\; \dots,\; t_{\nsamp-1} \}$ from an experiment on a dynamical system $\So$.  The data-generating system $\So$ is assumed to have the continuous-time state-space representation
\begin{subequations}
\label{eq:data_generating}
\begin{align}
 \dot x(t) &= f(x(t), u(t)) \\
 \ymodel(t)  &= g(x(t)),
\end{align}
\end{subequations}
where $x(t) \in \mathbb{R}^{\nx}$ is the system state at time $t$; $\dot x(t)$ denotes the  time derivative of $x(t)$; $\ymodel(t) \in \mathbb{R}^{\ny}$ is the noise-free output; $u(t) \in \mathbb{R}^{\nin}$ is the system  input; $f(\cdot, \cdot): \mathbb{R}^{\nx} \times \mathbb{R}^{\nin} \rightarrow \mathbb{R}^{\nx}$ and $g(\cdot): \mathbb{R}^{\nx} \rightarrow \mathbb{R}^{\ny}$ are the state and output mappings, respectively.
The measured output $y_k$ at time instant $t_k$, $k=0,1,\dots,\nsamp -1$, is  corrupted by a zero-mean additive noise $\eta_k$, \emph{i.e.}, $y_k =  \ymodel(t_k)  + \eta_k$. We assume that the input $u(t)$ can be  reconstructed (or reasonably approximated) for all time instants $t \in [t_0\; t_{N-1}] \subset \mathbb{R}$ from the samples $U$. 

{Our objective is to estimate from $\Did$ a \emph{simulation} model $M$ of the unknown dynamical system $\So$, \emph{i.e.}, the model $M$, given the input sequence $u(t)$, should generate an output sequence $\simul{y}(t)$ close to the true system output $\ymodel(t)$.\footnote{{We are not considering in this work \emph{prediction} models, which may also exploit previous 
\emph{measured} output values to predict $\ymodel(t)$}.}
}

{To be of practical use, the model should \emph{generalize} to conditions that are not encountered in the training dataset. For this reason, we shall evaluate the model performance in terms of the ``discrepancy''  between $y(t)$ and $\simul{y}(t)$ on an independent test dataset, where the system \eqref{eq:data_generating} is driven by an input sequence $u(t)$ different from the one used for training. Specifically, metrics such as the Root Mean Squared Error (RMSE) and the $R^2$ index shall be considered.}

{The dynamic nature of the system at hand calls for specialized models and learning techniques that can deal with the inherently sequential nature of the training sequences. Furthermore, these techniques have to be tailored for the considered simulation objective}. In this paper, we introduce flexible neural model structures {$\mathcal{M}$} that are suitable to represent generic dynamical systems as~\eqref{eq:data_generating}, allowing the modeler to embed domain knowledge to various degrees and to exploit neural networks to describe unknown model components.
Furthermore, we present fitting criteria and algorithms to train these neural dynamical model structures, {namely to select a model $M$ within the model structure $\mathcal{M}$ with good simulation performance, using the training dataset $\Did$.}
 
Overall, we aim at fitting a neural network model that exploits the dynamic constraints and different forms of available prior  knowledge of the data-generating system \eqref{eq:data_generating}.

\section{Neural dynamical models}
\label{sec:model_structure}
Let us consider a  \textit{model structure} $\MS = \{\M(\theta),\; \theta \in \mathbb{R}^{\ntheta}\}$, where $M(\theta)$ represents a dynamical model parametrized by a real-valued vector $\theta$.  We refer to \textit{neural model structures} as structures $\MS$ where some components of the model $M(\theta)$ are described by neural networks.
In the following, we introduce possible neural model structures  $\MS$ for dynamical systems. 

\subsection{General state-space model}
A general \emph{state-space neural  model structure} has form 
\begin{subequations}
\label{eq:neural_ss_general}
\begin{align}
 \dot x &= \NN_{\!f}(x, u; \theta) \label{eq:neural_ss_general_X}\\
 \ymodel &= \NN_{\!g}(x; \theta),
\end{align}
\end{subequations}
where $\NN_{\!f}$ and $\NN_{\!g}$ are {fully connected}, feedforward neural networks \cite{schmidhuber2015deep} of compatible size parametrized by $\theta \in \mathbb{R}^{\ntheta}$. For notational convenience, the time dependence of all signals  in~\eqref{eq:neural_ss_general}  is omitted.\footnote{In the rest of the paper, the time dependence of signals will be specified only when necessary.}

 {For the sake of exposition, in the case of neural networks $\NN_f$ and $\NN_g$ with a single hidden layer, the  network $\NN_f$ is characterized by an input layer with $\nx + \nin$ units corresponding to the system state $x$ and system input $u$; a certain number of hidden layers; and a linear output layer containing $n_x$ units corresponding to state derivatives. The  network $\NN_g$ has an input layer with $\nx$ units corresponding to the states $x$; a certain number of hidden layers; and a linear output layer with $\ny$ units corresponding to the system outputs. Thus, the single-hidden-layer neural networks $\NN_f$ and $\NN_g$ have structure
\begin{subequations}
\label{eq:FCFNN}
\begin{align}
 \NN_f(x,u;\theta) &= \mathcal{W}_{f,2}\left(\varepsilon(\mathcal{W}_{f,1} [x\; u]^\top + \beta_{f,1}) + \beta_{f,2}\right)\\
 \NN_g(x;\theta) &= \mathcal{W}_{g,2}\left(\varepsilon(\mathcal{W}_{g,1} x +  \beta_{g,1}) + \beta_{g,2} \right).
 \end{align}
\end{subequations}
In \eqref{eq:FCFNN}, the weight matrices $\mathcal{W}_{f,2}$, $\mathcal{W}_{f,1}$, $\mathcal{W}_{g,2}$, $\mathcal{W}_{g,1}$, and bias vectors $\beta_{f,2}$, $\beta_{f,1}$, $\beta_{g,2}$, $\beta_{g,1}$ are the tunable model parameters collectively represented in the parameter vector $\theta \in \mathbb{R}^{\ntheta}$ for notational compactness, while $\varepsilon(\cdot)$ is the neural network's non-linear \emph{activation function}, which is applied \emph{element-wise} to each element of its (matrix) input, producing an output of the same dimensionality.
Note that the number of rows of $\mathcal{W}_{f,1}$ must be equal to the number of columns of 
$\mathcal{W}_{f,2}$ (and to the number of elements in the vector $\beta_{f,1}$) and corresponds to the number of hidden units in the (single) hidden layer of $\NN_f$. Similar dimensionality constraints apply to the weight matrices and bias terms of the neural network $\mathcal{N}_g$.
}


{In the following, fully connected feedforward neural networks are always denoted with the letter
$\NN$ (using different subscripts to distinguish among them) and referred to in short as ``neural networks''. The number of input and output layers of these neural network should be clear to the reader from context. The exact network structure (number of hidden layers, number of hidden units per layer, type of activation function) is only specified for the examples in Section \ref{sec:example} and not in the main sections of the paper, as it is not required in the general discussion.}

The general neural model structure~\eqref{eq:neural_ss_general} can be tailored for the identification task at hand. Examples are illustrated in the remainder of this section. 

\subsection{Incremental model}
If a linear approximation of the system is available, an appropriate model structure could be
\begin{subequations}
\label{eq:neural_ss_lin}
\begin{align}
 \dot x   &= A_L x + B_L u + \NN_{\!f}(x, u; \theta)  \label{eq:neural_ss_lin_state}\\
 \ymodel   &= C_L x + \NN_{\!g}(x; \theta),
\end{align}
\end{subequations}
where $A_L$, $B_L$, and $C_L$ are matrices of compatible size describing the linear system approximation. For example, the values of these matrices can be estimated from the available training dataset $\Did$ using well-established  algorithms for linear system identification~\cite{ljung:1999system,van1994n4sid}.  
  Although model \eqref{eq:neural_ss_lin} is not more general than \eqref{eq:neural_ss_general}, it could be easier to train  as the neural networks $\NN_{\!f}$ and $\NN_{\!g}$ are supposed to capture only residual (nonlinear) dynamics.

\subsection{Fully-observed state model}
If the system state is known to be fully observed, the most  convenient representation is 
\begin{subequations}
\label{eq:neural_ss_full}
\begin{align}
 \dot x &= \NN_{\!f}(x, u; \theta) \label{eq:neural_ss_state} \\ 
 \ymodel     &= x,
\end{align}
\end{subequations}
where only the state mapping neural network $\NN_{\!f}$ is learned, while the output mapping is fixed to identity.

\subsection{Physics-based model}
Tailor-made architectures could  be used to embed specific physical knowledge in the neural model structure.

\begin{figure}
\centering
 \includegraphics[width=.4\textwidth]{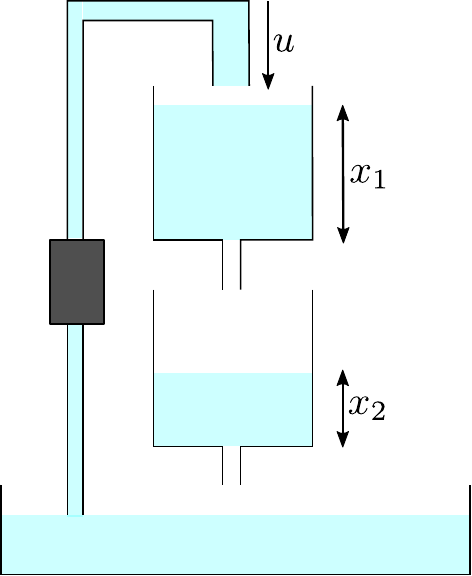}
 \caption{Schematics of the cascaded two-tank system.}
 \label{fig:CTS}
\end{figure}

For instance, let us consider the \emph{Cascaded Tanks System} (CTS) schematized in Figure~\ref{fig:CTS}. 
The CTS is a fluid level control system consisting of two tanks with free outlets fed by a pump.  
Water is pumped from a bottom reservoir into the upper tank by a controlled pump.
The  water in the upper tank flows through a small opening into the lower tank, and from another small opening from the lower tank to the reservoir. 

The system input $u$ is the water flow from the bottom reservoir feeding the upper tank.
The state variables $x_1$ and $x_2$ are the water level in the upper and lower tanks, respectively.\footnote{With some abuse of notation, the subscripts in $x_1$ and $x_2$ simply denote the variable name and not a time index. The same notation will be used for the examples in Section \ref{sec:example}.}
As in the illustrative  example further discussed in Section~\ref{sec:example}, we consider the case where only the second state $x_2$ is  measured.

The following dynamical model for the CTS can be derived from conservation laws and Bernoulli's principle \cite{schoukens2017three}:
\begin{subequations}
\label{eq:CTS}
\begin{align}
 \dot x_1 &= -k_1 \sqrt x_1 + k_4 u\\
 \dot x_2 &= k_2\sqrt x_1 - k_3 \sqrt{x_2}\\
      y   &= x_2.
\end{align}
\end{subequations}
Based on this physical knowledge, an appropriate neural dynamical model for the CTS is
\begin{subequations}
\label{eq:neural_CTS}
\begin{align}
 \dot x_1 &= \NN_{\!f_1}(x_1, u) \\
 \dot x_2 &= \NN_{\!f_2}(x_1, x_2) \\
      y   &= x_2
\end{align}
\end{subequations}
capturing the information that: ($i$) the system dynamics can be described by a two-dimensional state-space model; ($ii$) the state $x_2$ is measured; ($iii$) the  state $x_1$ does not depend on $x_2$; and $(iv)$ the state $x_2$  does not depend directly on $u$, given the current value of $x_1$.
This neural model takes advantage of the available process knowledge, while leaving representational capabilities to describe unmodeled effects such as  fluid viscosity,  nonlinearities of the actuators, and water overflow that may happen when the tanks are completely filled. 

Embedding physical knowledge in neural model structures is a very active and promising trend in deep learning \citep{raissi2019physics}.
For instance, recent contributions propose specialized structures that are suitable to describe systems satisfying general physical principles such a energy conservation.
In these cases, a physics-based neural network may be used to learn the system's Hamiltonian or Lagrangian function, instead of the individual components its ODE representation as independent terms \citep{greydanus2019hamiltonian, lutter2019deep}. 

In general, including domain knowledge in the model structure is useful to restrict the search space, while leaving enough representation capacity.
Therefore, a better generalization performance is expected from the trained models. Furthermore, the estimated models exactly satisfy (by design) the prior assumptions and thus they are generally easier to diagnose, interpret and exploit for advanced tasks such as state estimation, fault detection and closed-loop control.


\section{Training neural dynamical models}
\label{sec:training}

In this section, we present algorithms aimed at fitting the model structures  introduced in Section \ref{sec:model_structure} to the training dataset $\Did$. 

{
\subsection{Simulation error minimization}
}
For \emph{fixed} values of neural network parameters $\theta$, for \emph{given} initial condition $x_0 = x(0)$, and under the model  structure~\eqref{eq:neural_ss_general},  the open-loop state simulation $\simul{x}(t)$ is the solution of the Cauchy problem:
\begin{subequations}
\label{eq:cauchy}
\begin{align}
{\simul{ \dot x}}(t) &= \NN_{\!f}\big(\simul{x}(t), u(t);\; \theta \big) \\
 \simul{x}(0)        &= x_0,
\end{align}
\end{subequations}
and the simulated output ${\simul{\est{y}}}(t)$ is 
\begin{equation}
\label{eq:output_equation}
 {\simul{\est{y}}}(t;\;\theta, x_0) = \NN_{\!g}(\simul{x}(t;\;\theta, x_0);\;\theta). 
\end{equation}
Different  ODE solution schemes~\citep{quarteroni2010numerical}, may be applied to numerically solve problem~\eqref{eq:cauchy}.  
Formally, 
\begin{equation}
\label{eq:cauchy_approx}
 \simul{x}(t;\;\theta, x_0) = \text{ODEINT}\left(t;\; \NN_{\!f}\left(\cdot,\; \cdot;\; \theta\right), u(\cdot), x_0 \right)
\end{equation}
will represent the solution of the Cauchy problem~\eqref{eq:cauchy} obtained  using a numerical scheme of choice (explicit or implicit, single-step or multi-step, single-stage or multi-stage) denoted as ODEINT.


{
According to the simulation error minimization criterion, 
}
the neural network parameters $\theta$ can be obtained by minimizing the simulation error norm  
\begin{equation}
\label{eq:simulation_error_cost_function}
J(\theta, x_0) = \frac{1}{\nsamp} \sum_{k=0}^{N-1} \bigg|\bigg|
\overbrace{\simul{\est{y}}(t_k;\; \theta, x_0) -   y_k}^{e_k}
\bigg|\bigg|^2
\end{equation}
with respect to both the network parameters $\theta$ and the state initial condition $x_0$, with $\simul{\est{y}}$  defined by Equations 
\eqref{eq:output_equation}-\eqref{eq:cauchy_approx}.  



\begin{figure}
 \centering
 \includegraphics[width=.9\textwidth]{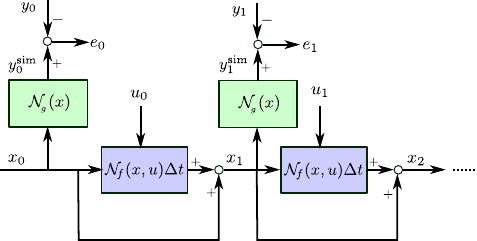}
 \caption{Computational graph associated with the forward Euler ODE integration scheme with constant step size $\Delta t$ applied to system \eqref{eq:neural_ss_general}.
  Measurements are assumed to be equally spaced with the same constant rate $\Delta t$.}
 \label{fig:forward_euler_graph}
\end{figure}

From an implementation perspective, when ODEINT is an explicit ODE solver, the full computational graph producing the cost function \eqref{eq:simulation_error_cost_function} can be constructed using standard differentiable blocks. For example, Figure~\ref{fig:forward_euler_graph} represents the computational graph obtained by applying a forward Euler scheme with constant step size $\Delta t$, assuming that the measurements in $\Did$ are collected at the same rate $\Delta t$.
{It is interesting to note the similarity between this computational graph and the one of the \emph{residual network} structure~\cite{he2016deep}.}
{Note that number of repeated blocks in the computational graph (thus, the depth of the architecture) corresponds to the length of the training sequence.}

Therefore, in the case  of an explicit ODE solver, the derivatives of the loss $J(\theta, x_0)$ with respect to $\theta$ and $x_0$ 
required for gradient-based minimization can be obtained using standard back-propagation through the elementary solver steps. 
Thus, a procedure minimizing $J(\theta, x_0)$ can be implemented using available deep learning software. 

Recently, an alternative approach to differentiate through the ODE solution based on backward time integration of \emph{adjoint sensitivities} has been proposed \cite{chen2018neural}. Following this approach, implicit ODE solvers may be  adopted as well. 

In either case, from a computational perspective,  simulating over time has an intrinsically sequential nature and offers scarce opportunities for parallelization. Thus, in practice,  minimizing the simulation error with a gradient-based method over the entire training dataset  $\Did$ may be inconvenient or even unfeasible in terms of memory allocation and computational time. 

{Moreover, as thoroughly analyzed in \cite{ribeiro2020smoothness}, numerical minimization of the 
simulation-error norm may be hard as its Lipschitz constant might blow up exponentially with the simulation length for certain system having for instance unstable or chaotic dynamics.}

\begin{remark}
In many cases, optimizing the initial condition $x_0$ is not necessary in simulation error minimization as this quantity may be available from physical considerations. For instance, in the cascaded tanks system presented above, the initial values of the tank levels may be known.  Even when the initial condition is unknown, its effect may be negligible, as in the case for measurements collected from a \emph{fading memory} system $\So$ on a sufficiently long time horizon $[t_0 \; t_{N-1}]$.
\end{remark}

{
\begin{remark}
As in all non-convex optimization problems, a gradient-based optimization algorithm is not guaranteed to reach a global minimizer of the simulation error loss. This prevents us from deriving formal proofs of convergence for gradient-based simulation error minimization, as well as for all other algorithms presented in this paper. Nonetheless, this fundamental theoretical limitation does not seem to constitute a severe limitation for deep learning applications in practice. As discussed in~\cite[Ch.~8.2]{goodfellow2016deep}, among the stationary points of typical loss functions involving deep neural networks, local minimizers with a ``high'' cost (w.r.t. the global minimum of the loss) are rather uncommon compared, for instance, to saddle points. Empirically, gradient descent seems to be able to escape from these saddle points in many cases.
\end{remark}
}
\subsection{Truncated simulation error minimization}
In order to reduce the computational burden {and circumvent the numerical issues possibly affecting} full simulation error minimization approach previously discussed,  
the simulated output $\simul{\est{y}}(t;\;\theta, x_0)$ can be obtained by simulating the system state $\simul{x}(t;\; \theta, x_0)$ in~\eqref{eq:cauchy_approx} on several smaller portions of the training set $\Did$.  

 For efficient implementation, the \emph{truncated} simulation error minimization algorithm presented in this section  processes \emph{batches} containing $\batchsize$ subsequences extracted from $\Did$. In principle, the simulations can be carried out simultaneously for all the subsequences in the batch by exploiting parallel computing. 

\begin{figure}
\centering
\includegraphics[width=.9\textwidth]{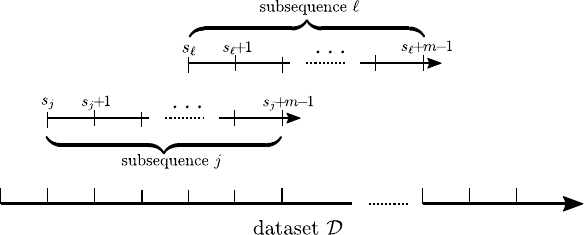}
\caption{Representation of two subsequences of length $m$ extracted from the training dataset  $\mathcal{D}$.  $s_j$ and $s_\ell$ represent the starting indexes of subsequences $j$ and $\ell$, respectively. }
\label{fig:subsequences}
\end{figure}

A batch is completely specified by a \emph{batch starting index vector} ${s} \in \mathbb{N}^{\batchsize}$ defining the initial sample of each subsequence and a \emph{sequence duration} ${\seqlen} \in \mathbb{N}$ defining the number of  samples contained in each subsequence, where each element $s_j$ of $s$ satisfies $s_j\leq \nsamp - \seqlen - 1$. Thus, for instance, the $j$-th output subsequence in a batch contains the measured output samples $\{y_{s_j},$ $y_{s_j+1},$ $\dots, y_{s_j+\seqlen-1}\}$ 
(see Figure \ref{fig:subsequences} for graphical representation).

For notational convenience, we can arrange the batch data in the following \emph{tensors}:
\begin{itemize}
 \item output $\tens{y} \in \mathbb{R}^{\batchsize \times \seqlen \times n_y}$, where $\tens{y}_{j,h} = {y_{s_j+h}}$
 \item input $\tens{u} \in \mathbb{R}^{\batchsize \times \seqlen \times n_u}$, where $\tens{u}_{j,h}
 = {u_{s_j+h}}$
\item relative time $\tens{\tau} \in \mathbb{R}^{\batchsize \times \seqlen}$, where $\tens{\tau}_{j,h} = {t_{s_j+h} - t_{s_j}}$,
\end{itemize}
for $j=0,1,\dots,\batchsize-1$ and $h=0,1,\dots, \seqlen-1$. Furthermore, we define the tensors
\begin{itemize}
  \item  simulated  state $\simul{\tens{x}} \in \mathbb{R}^{\batchsize \times \seqlen \times n_x}$
\item simulated output $\simul{\est{{\tens{y}}}} \in \mathbb{R}^{\batchsize \times \seqlen \times n_y}$,
\end{itemize}
that will be used to store simulated state and output values at the corresponding batch time instants.

Note that the third dimension of the output tensor $\tens{y} \in \mathbb{R}^{\batchsize \times \seqlen \times n_y}$ corresponds to the output channel. With some abuse of notation, the tensor $\tens{y}$ is addressed by only two indexes  because we do not need to specify a particular output channel. The same notation is used for tensors $\tens{u}$, $\simul{\tens{x}}$, 
and $\simul{\est{{\tens{y}}}}$.  

For the subsequences $ j=0,1,\dots,\batchsize-1$, the state evolution  in the relative time intervals $[0\; \;  \tens{\tau}_{j, \seqlen-1}]$ (which corresponds to the absolute time interval $[t_{s_j}\; t_{s_j+\seqlen-1}]$) is the solution of the Cauchy problem 
\begin{align} \label{eqn:TS:Cauchy}
\simul{{\dot x}}_{j}(\tau) &= \NN_{\!f}(\simul{{x}}_{j}(\tau), u(t_{s_j} + \tau);\;\theta)
\end{align}
for a given initial condition $\simul{{x}}_{j}(0)$. The solution of the Cauchy problem~\eqref{eqn:TS:Cauchy} 
is numerically  approximated by 
\begin{subequations}
\begin{align} \label{eqn_ODEint}
\simul{{x}}_{j}(\tau) &= \text{ODEINT}(\tau;\; \NN_{\!f}(\cdot, \cdot;\; \theta), u(\cdot), \simul{{x}}_{j}(0)),
\end{align}
and the simulated output for the $j$-th subsequence is then given by
\begin{align} \label{eqn:xsim}
    \simul{\est{{y}}}_j(\tau) &= \NN_{g}( \simul{{x}}_{j}(\tau);\; \theta).
\end{align}
\end{subequations}
We can now arrange the simulated state and output at the measurement time instants in the tensors 
$\simul{\est{\tens{y}}}$ and $\simul{\tens{x}}$, respectively:
\begin{subequations}
\begin{align}
 \simul{\est{\tens{y}}}_{j,h} &= \simul{\est y}_j(\tens{\tau}_{j,h})\\
 \simul{{\tens{x}}}_{j,h}     &= \simul{x}_j(\tens{\tau}_{j,h}),
 \end{align}
\end{subequations}
for $j=0,1,\dots,\batchsize-1$ and $h=0,1,\dots,\seqlen-1$. 

As opposed to the full simulation error minimization case, the choice of appropriate initial conditions $\simul{{x}}_{j}(0)$ in the Cauchy problem~\eqref{eqn:TS:Cauchy} is here critical. Indeed, the effect of these initial conditions cannot in general be neglected as the duration of a subsequence is typically much shorter than the whole dataset $\Did$.
Furthermore, the system state is unlikely to be known \emph{a priori} at all different time instants. 
For this reason, we introduce  an additional  set of free    variables   $\hidden{X} = \{\hidden{x}_0,\hidden{x}_1,\dots,\hidden{x}_{\nsamp-1}\}$,  where $\hidden{x}_k \in \mathbb{R}^{\nx}$  represents the (unmeasured) system 
state at the  measurement time $t_k$. 
For a given batch, the set of subsequences' initial conditions   $\simul{{x}}_{j}(0)$  is then constructed as $\simul{{x}}_{j}(0) = \hidden{x}_{s_j}$ (with $j=0,\ldots,\batchsize-1$) and optimized along with the neural network parameters $\theta$ in order to minimize the fitting cost 
\begin{equation}
\label{eq:simulation_cost_fit}
J_{\rm fit}(\theta,\hidden{X}) = \frac{1}{\batchsize \seqlen}\sum_{j=0}^{\batchsize-1}\sum_{h=0}^{\seqlen-1} \norm{\est{\tens{y}}^{\rm sim}_{j,h}(\theta, \hidden{X}) - {{\tens{y}}}_{j,h}}^2. 
\end{equation}
It is important to remark that the simulated output $\est{\tens{y}}^{\rm sim}_{j,h}(\theta, \hidden{X})$ is a function  of both the neural network parameters $\theta$ and the initial conditions $\hidden{x}_{s_j} \in  \hidden{X}$. 

 Since the fitting cost $J_{\rm fit}(\theta,\hidden{X})$ defined above has $\nsamp$ additional degrees of freedom w.r.t. the full simulation error minimization case, a price could be paid in terms of lack of generalization of the estimated model.


In order to reduce the degrees of freedom in the minimization  of the loss $J_{\rm fit}$, 
the variable  $\hidden{X}$ used to construct the  initial conditions for the Cauchy problem~\eqref{eqn:TS:Cauchy} can be enforced to  represent the unknown system state and thus to be \emph{consistent} with
the neural model structure~\eqref{eq:neural_ss_general_X} (namely, the state sequence  $\hidden{X}$ should satisfy the neural ODE equation~\eqref{eq:neural_ss_general_X}). To this aim, we introduce a regularization term $J_{\rm reg}(\theta, \hidden{X})$  penalizing the distance between the  state ${\tens{x}}^{\mathrm{sim}}$ (simulated through~\eqref{eqn:xsim}) and the optimization variables in $\hidden{X}$. Specifically, the regularization term $J_{\rm reg}(\theta, \hidden{X})$  is defined as 
 \begin{equation}
 \label{eq:simulation_cost_regularizer}
 J_{\rm reg}(\theta,\hidden{X}) = \frac{1}{\batchsize \seqlen}\sum_{j=0}^{\batchsize-1} \sum_{h=0}^{\seqlen-1} \norm{{\tens{x}}^{\rm sim}_{j,h}(\theta, \hidden{X}) - \hidden{{\tens{x}}}_{j,h}(\hidden{X})}^2, 
 \end{equation}
where 
 $\hidden{\tens{x}}$  is a tensor with the same size and structure as ${\tens{x}}$, but populated with  samples from $\hidden{X}$, \emph{i.e.},
 ${\hidden{\tens{x}}}_{j,h} = \hidden{x}_{s_{j}+h}$.

The overall loss is then constructed as a weighted sum of the two objectives, namely:
\begin{align} \label{eqn:Jtot}
J_{\rm tot }(\theta,\hidden{X}) = J_{\rm fit}(\theta,\hidden{X}) +\alpha J_{\rm reg}(\theta, \hidden{X})
\end{align}
with regularization weight $\alpha \geq 0$.

\begin{algorithm}
	\caption{Truncated simulation error minimization}
	\label{algo:minibatch_multistep_simulation}
\small
	\textbf{Inputs}: training dataset $\Did$; number of iterations $n$; batch size $\batchsize$; length of subsequences~$\seqlen$; learning rate $\lambda>0$;  regularization weight $\alpha \geq 0$.
\vspace*{-.0cm}\hrule\vspace*{.0cm}
	\begin{enumerate}[label=\arabic*., ref=\theenumi{}]
        \item  \textbf{initialize} the neural network parameters $\theta$ and the hidden  state sequence  $\hidden{X}$; 
		\item  \textbf{for} $i=0,\ldots,\numiter-1$ \textbf{do}
		\begin{enumerate}[label=\theenumi{}.\arabic*., ref=\theenumi{}.\theenumii{}]
			\item \textbf{select} batch start index  vector $s \in \mathbb{N}^q$;
			\item \textbf{populate} tensors
			\begin{align*}
			& {{\tens{y}}}_{j,h}=y_{s_j+h}, \qquad
			 \hidden{{\tens{x}}}_{j,h}=\hidden{x}_{s_j+h}, \qquad
			 {{\tens{\tau}}}_{j,h}={t_{s_j+h} - t_{s_j}}\\
			 & \textrm{for  } j\!=\!0,1,\dots,\batchsize\!-\!1 \textrm{\ \ and\ \ } h\!=\!0,1,\dots,\seqlen\!-\!1
			\end{align*}
			and set of initial conditions
			\begin{equation*}
			{x}^{\mathrm{sim}}_{j}(0) = \hidden{x}_{s_j}, 
			\textrm{\; for }j\!=\!0,1,\dots,\batchsize\!-\!1;
			\end{equation*}
			\item  \textbf{simulate} state and output   
 			\begin{align*} 
 			\simul{{x}}_{j}(\tau) &= \text{ODEINT}(\tau;\; \NN_{\!f}(\cdot, \cdot; \theta), {x}^{\mathrm{sim}}_{j}(0) )  \\ 	
 {{y}}_{j}^{\mathrm{sim}}(\tau)  &= \NN_{g}(\simul{{x}}_{j}(\tau);\theta) 
\end{align*} 
 for $j\!=\!0,1,\dots,\batchsize\!-\!1 \textrm{\ \ and\ \ }  \tau \in [0\; \tau_{s_{j+m}}]$;
\item  \textbf{populate} tensors $\simul{\est{\tens{x}}}$ and $\simul{{\tens{y}}}$ as   
\begin{align*}
 \simul{{\tens{x}}}_{j,h} = \simul{x}_j(\tens{\tau}_{j,h}) \\
  \simul{\est{\tens{y}}}_{j,h} = \simul{\est y}_j(\tens{\tau}_{j,h}) 
 \end{align*}
 for $j\!=\!0,1,\dots,\batchsize\!-\!1$ and $h\!=\!0,1,\dots,\seqlen\!-\!1$;
			\item \textbf{compute} the cost  
\begin{multline*}
\label{eq:simulation_cost_tot_algo}
J_{\rm tot}(\theta,\hidden{X}) = 
\overbrace{ \frac{1}{\batchsize \seqlen}\sum_{j=0}^{\batchsize-1}\sum_{h=0}^{\seqlen-1} \norm{{\tens{\est{y}}}^{\rm sim}_{j,h}(\theta, \hidden{X}) - {{\tens{y}}}_{j,h}}^2}^{J_{\rm fit}} + \\
+\alpha\overbrace{\frac{1}{\batchsize \seqlen} \sum_{h=0}^{\seqlen-1} \norm{{\tens{x}}^{\rm sim}_{j,h}(\theta, \hidden{X}) - \hidden{{\tens{x}}}_{j,h}(\hidden{X})}^2}^{J_{\rm reg}};
\end{multline*}

\item \textbf{evaluate} the gradients $\nabla_\theta J_{\rm tot}=\frac{\partial J_{\rm tot}}{\partial \theta}$ and  
			$\nabla_{\hidden{X}} J_{\rm tot}=\frac{\partial J_{\rm tot}}{\partial \hidden{X}}$ at the current values of $\theta$ and $\hidden{X}$;
			\item  \textbf{update} optimization variables $\theta$ and $\hidden{X}$:
			\begin{equation*}
			\label{eq:SGD}
			\begin{split}
			 \theta &\leftarrow \theta - \lambda \nabla_\theta J_{\rm tot}  \\
			 \hidden{X} & \leftarrow \hidden{X} - \lambda \nabla_{\hidden{X}} J_{\rm tot};
			\end{split}
			\end{equation*}
		\end{enumerate}
	\end{enumerate}
	\vspace*{-.0cm}\hrule\vspace*{.1cm}
	\textbf{Output}:  neural network parameters $\theta$. 
\end{algorithm} 

 Algorithm \ref{algo:minibatch_multistep_simulation} details   the steps  required by the proposed truncated simulation error method   
 to train  a dynamical neural model via gradient-based optimization. 
 
 In Step 1, the neural network parameters $\theta$ and the ``hidden'' state variables in $\hidden{X}$ are initialized. For instance, the initial values of  $\theta$ and  $\hidden{X}$ can be set to small random numbers. Alternatively, if a measurement/estimate of some of the system's state variables is available, it can be used to initialize certain  components of $\hidden{X}$.

   Then, at each iteration $i=0,\ldots,n-1$ of the gradient-based training algorithm, the following steps are executed.   
Firstly, the batch start vector $s \in \mathbb{N}^\batchsize$ is selected with $s_j \in [0 \ \  \nsamp-\seqlen-1],\; j=0,1,\dots,\batchsize-1$ (Step 2.1). The indexes in $s$ can be either (pseudo)randomly generated, or chosen deterministically.\footnote{For an efficient use of the training data $\Did$, $s$ has to be chosen in such a way that all samples are visited with equal frequency  during the  iterations of Algorithm \ref{algo:minibatch_multistep_simulation}.}  
Then,  tensors ${{\tens{y}}}$, ${\hidden{\tens{x}}}$,  and ${{\tens{\tau}}}$ are populated with the corresponding values in $\Did$ and $\hidden{X}$. The initial conditions ${x}^{\mathrm{sim}}_{j}(0)$, with $j=0,1,\dots,\batchsize-1$ are also obtained from $\hidden{X}$  (Step 2.2). 
Subsequently, for each subsequence $j$, the system state and output are simulated within the time interval $[0\; \tau_{s_{j+\seqlen - 1}}]$, starting from the initial condition ${x}^{\mathrm{sim}}_{j}(0)$  (Step 2.3). Then, the values of the simulated state and output at the (relative) measurement times $\tens{\tau}_{j,h}$ (with $h=0,\ldots,\seqlen-1$) are collected in tensors $\simul{\est{\tens{x}}}$ and $\simul{{\tens{y}}}$ (Step 2.4), and  used to  construct the loss $J_{\rm tot}(\theta,\hidden{X})$  (Step 2.5).
Next, the gradients of the cost with respect to the optimization variables $\theta$, $\hidden{X}$ are obtained in (Step 2.6), either by back-propagation through the solver steps or exploiting the adjoint sensitivity method suggested in \cite{chen2018neural}.
Lastly, the optimization variables are updated via gradient descent with \emph{learning rate} $\lambda$ (Step 2.7). Improved variants of  gradient descent  such as \emph{Adam}~\citep{kingma2014adam} can be alternatively adopted at Step 2.7. 

Note that, at each iteration $i$ of the gradient descent algorithm, the cost $J_{\rm tot}(\theta,\hidden{X})$ may depend on just a subset of hidden state variables $\hidden{X}$. 
Thus, only for those components at each iteration the gradient vector $\nabla_{\hidden{X}} J_{\rm tot}=\frac{\partial J_{\rm tot}}{\partial \hidden{X}}$ is non-zero and an update is performed.

\begin{remark}
The computational cost of evaluating  the gradient of $J_{\rm tot}$ is proportional to the number of solver steps executed in~\eqref{eqn_ODEint}. In the case the solver step is equal to the sampling intervals $t_{k+1}-t_k$ (which corresponds to $\seqlen$ solver steps),    running truncated  simulation error minimization with $\seqlen \ll \nsamp$  is thus significantly faster than full simulation error minimization.  
Furthermore, the computations for the $\batchsize$ subsequences can be carried out independently, and thus  parallel computing can be exploited. 
\end{remark}
 
 \begin{remark}
The use of a regularization term enforcing consistency of the hidden variable $\hidden{X}$ with the system dynamics has analogies with the \emph{multiple shooting} numerical method for optimal control of dynamical systems \cite{bock1984multiple}. As in multiple shooting, the original optimization problem defined over a long time window is split into smaller intervals, which are then processed 
independently. While in multiple shooting the intervals are fixed and continuity of the solution is enforced by equality constraints at the interval boundaries, in our approach the subsequences may have different starting points, and the regularization term promoting consistency of the hidden variable $\hidden{X}$ is defined for all time steps.
\end{remark}

\begin{remark}
{
Our regularized truncated simulation error minimization criterion \eqref{eqn:Jtot} has also similarities with the cost function minimized by the neural moving horizon estimator presented in \cite{alessandri2011moving}. Furthermore, a by-product of our identification algorithm is a state estimate $\hidden{X}$ for all the time instants in the training dataset that may be interpreted,
for each time instant, as an average of all $m$-length moving horizon estimators including that time instant, based on the learned model dynamics. 
Note that while in \cite{alessandri2011moving} the true system dynamics is assumed to be known and a neural network is used to model the state estimation function, in our approach the true dynamics is unknown and jointly approximated as neural networks together with the hidden state sequence $\hidden{X}$, by minimizing the regularized fitting criterion~\eqref{eqn:Jtot}.
}
\end{remark}

 In the next paragraph, we introduce an alternative method for fitting neural dynamical models that does not require simulation through time, and thus is even more convenient from a computational perspective, allowing full parallelization at time step level. 

\subsection{Soft-constrained integration}
The optimization variables $\hidden{X}$ previously introduced  for truncated  simulation error minimization are regularized to be consistent with the fitted system dynamics through the cost $J_{\rm reg}$~\eqref{eq:simulation_cost_regularizer}. 
Therefore, $\hidden{X}$ implicitly provides another estimate of the unknown system state,
and thus of the output $Y$ for given $\NN_{g}$. This estimate can be compared with the measured samples $Y$ to define an alternative fitting objective. 
This suggests the following variant for the fitting term $J_{\rm fit}$: 
\begin{equation}
\label{eq:consistency_cost_fit}
J_{\rm fit}(\hidden{X}) = \frac{1}{\batchsize \seqlen}\sum_{j=0}^{\batchsize-1}\sum_{h=0}^{\seqlen-1} \norm{{\hidden{\tens{y}}}_{j,h} - {{\tens{y}}}_{j,h}}^2,
\end{equation}
where ${\hidden{\tens{y}}_{j,h}} = \NN_{\!g}({\hidden{\tens{x}}_{j,h}})$. 

Experimentally, using \eqref{eq:consistency_cost_fit} instead of $\eqref{eq:simulation_cost_fit}$ as fitting term does not significantly alter the properties of truncated simulation error minimization.
Furthermore, the wall-clock execution time of Algorithm~\ref{algo:minibatch_multistep_simulation} with the modified fitting term \eqref{eq:consistency_cost_fit} is still
 dominated by the $m$-step simulation (and back-propagation) still required to compute the gradient of  the regularization term $J_{\rm reg}$ in \eqref{eq:simulation_cost_regularizer}. 
In order to formulate a faster learning algorithm, an alternative 
regularization term $J_{\rm reg}$ promoting consistency of the hidden state variables, but not requiring time simulation should be devised.

 The consistency-promoting regularizer $J_{\rm reg}$ considered in the soft-constrained integration method penalizes the violation of a numerical ODE integration scheme applied to the hidden state variables $\hidden{X}$, independently at each integration step. 
For instance, the forward Euler scheme can be enforced by means of the regularization term 
\begin{equation}
\label{eq:soft_forward_euler}
J_{\rm reg}(\hidden{X}, \theta) =  \sum_{j=0}^{\batchsize-1}\sum_{h=1}^{\seqlen-1} \norm{ \hidden{\tens{x}}_{j,h} -
\hidden{\tens{x}}_{j,h-1} - \Delta t \NN_{\!f}(\hidden{\tens{x}}_{j,h-1}, \tens{u}_{j,h-1})}^2, 
\end{equation}
assuming for notation simplicity a constant step size $\Delta t$.
If the regularization term \eqref{eq:soft_forward_euler} is reduced to a ``small value'' through optimization, then the  hidden variables $\hidden{X}$ will (approximately) satisfy the forward Euler scheme.

\begin{algorithm}
	\caption{Soft-constrained integration  for the forward Euler scheme}
	\label{algo:minibatch_consistency_method}
\small
	\textbf{Inputs}: training dataset $\Did$; number of iterations $n$; batch size $\batchsize$; length of subsequences~$\seqlen$; learning rate $\lambda>0$;  regularization weight $\alpha \geq 0$.
\vspace*{-.0cm}\hrule\vspace*{.0cm}
	\begin{enumerate}[label=\arabic*., ref=\theenumi{}]
        \item  \textbf{initialize} the neural network parameters $\theta$    and the hidden state sequence $\hidden{X}$; 
		\item  \textbf{for} $i=0,\ldots,\numiter-1$ \textbf{do}
		\begin{enumerate}[label=\theenumi{}.\arabic*., ref=\theenumi{}.\theenumii{}]
			\item \textbf{select} batch start index  vector $s \in \mathbb{N}^q$;
			\item \textbf{populate} tensors
			\begin{align*}
			& {{\tens{y}}}_{j,h}=y_{s_j+h}, \qquad
			 \hidden{{\tens{x}}}_{j,h}=\hidden{x}_{s_j+h}, \qquad
			 {{\tens{u}}}_{j,h}=u_{s_j+h}, \qquad \\
			 & \textrm{for  } j\!=\!0,1,\dots,\batchsize\!-\!1 \textrm{\ \ and\ \ } h\!=\!0,1,\dots,\seqlen\!-\!1;
			\end{align*}			
			\item \textbf{compute} the cost 
\begin{multline*}
J_{\rm tot}(\theta,\hidden{X}) = 
\overbrace{\frac{1}{\batchsize \seqlen}\sum_{j=0}^{\batchsize-1}\sum_{h=0}^{\seqlen-1} \norm{{\hidden{\tens{y}}}_{j,h} - {{\tens{y}}}_{j,h}}^2}^{J_{\rm fit}}  + \\
+\alpha \overbrace{\frac{1}{\batchsize \seqlen} \sum_{j=0}^{\batchsize-1}\sum_{h=1}^{\seqlen-1} \norm{ \hidden{\tens{x}}_{j,h} -
\hidden{\tens{x}}_{j,h-1} - \Delta t \NN_{\!f}(\hidden{\tens{x}}_{j,h-1}, \tens{u}_{j,h-1})}^2}^{J_{\rm reg}},
\end{multline*}
with $\hidden{\tens{y}}_{j,h} = \NN_{\!g}(\hidden{\tens{x}}_{j,h})$;

\item \textbf{evaluate} the gradients $\nabla_\theta J_{\rm tot}=\frac{\partial J_{tot}}{\partial \theta}$ and  
			$\nabla_{\hidden{X}} J_{\rm tot}=\frac{\partial J_{\batchsize,\seqlen}}{\partial \hidden{X}}$ at the current values of $\theta$ and $\hidden{X}$;
			\item  \textbf{update} optimization variables $\theta$ and $\hidden{X}$:
			\begin{equation*}
			\begin{split}
			 \theta &\leftarrow \theta - \lambda \nabla_\theta J_{\rm tot}  \\
			 \hidden{X} & \leftarrow \hidden{X} - \lambda \nabla_{\hidden{X}} J_{\rm tot};
			\end{split}
			\end{equation*}
		\end{enumerate}
	\end{enumerate}
	\vspace*{-.0cm}\hrule\vspace*{.1cm}
	\textbf{Output}:  neural network parameters $\theta$. 
\end{algorithm} 

Algorithm \ref{algo:minibatch_consistency_method} details the training steps   when the forward Euler numerical scheme is used to enforce consistency of the hidden state variables $\hidden{X}$ with the model dynamics. In Step 1, the neural network parameters $\theta$, and the sequence of hidden variables $\hidden{X}$ are initialized.  Then, at each iteration $i=0,1,\dots, \numiter-1$ of the gradient-based training algorithm, the following steps are executed.
Firstly, the batch start vector $s \in \mathbb{N}^\batchsize$ is selected with $s_j \in [0 \ \  \nsamp-\seqlen-1],\; j=0,1,\dots,\batchsize-1$ (Step 2.1)
and the tensors ${{\tens{y}}}$, ${\hidden{\tens{x}}}$, ${\hidden{\tens{u}}}$  are populated with the corresponding samples in $\Did$ (Step 2.2), similarly as in Algorithm \ref{algo:minibatch_multistep_simulation}
 
Then, the loss $J_{\rm tot}$  is computed (Step 2.3)  as a weighted sum of the fitting cost $J_{\rm fit}$ in~\eqref{eq:consistency_cost_fit} and the regularizer $J_{\rm reg}$~\eqref{eq:soft_forward_euler}. Note that, unlike Algorithm~\ref{algo:minibatch_multistep_simulation},   Algorithm~\ref{algo:minibatch_consistency_method} does not require $\seqlen$-step  simulation. 
The  gradients  of $J_{\rm tot}$ are obtained using standard back-propagation and used to perform a gradient-based minimization step (Steps 2.4 and 2.5).


The potential advantage of the proposed soft-constrained integration method over the truncated simulation error minimization is twofold.  Firstly, implicit integration schemes can be enforced with no additional computational burden with respect to explicit ones. For instance, the backward Euler integration scheme can be implemented simply by modifying the consistency term $J_{\rm reg}$ to  \begin{equation}
\label{eq:soft_backward_euler}
J_{\rm reg}(\hidden{X}, \theta) =  \sum_{j=0}^{\batchsize-1}\sum_{h=1}^{\seqlen-1} \norm{\hidden{\tens{x}}_{j,h} - 
\hidden{\tens{x}}_{j,h-1} - \Delta t \NN_{\!f}(\hidden{\tens{x}}_{j,h}, \tens{u}_{j,h})}^2,
\end{equation}
while the Crank-Nicolson scheme corresponds to
\begin{equation}
\label{eq:soft_crank-nicolson}
J_{\rm reg}(\hidden{X}, \theta)\! =\!  \sum_{j=0}^{\batchsize-1}\sum_{h=1}^{\seqlen-1} \! ||\hidden{\tens{x}}_{j,h}\! -\! 
\hidden{\tens{x}}_{j,h-1}\! -\! \frac{\Delta t}{2} \! \left( \NN_{\!f}(\hidden{\tens{x}}_{j,h}, \tens{u}_{j,h})\!+\!\NN_{\!f}(\hidden{\tens{x}}_{j,h-1}, \tens{u}_{j,h-1}) \right)||^2.
\end{equation}
Other implicit schemes such as the multi-step Backward Differentiation Formula (BDF) and Adams-Moulton (AM), or multi-stage implicit Runge Kutta (RK) methods may be similarly implemented, leading to a potential increase in the accuracy of the ODE numerical solution \cite{quarteroni2010numerical}. 
Secondly, the resulting cost function splits up as a sum of independent contributions for each time step, thus enabling the fully parallel implementation of gradient-based optimization. This leads to significant computational advantages and a reduced wall-clock execution time. 

On the other hand, in the proposed soft-constrained integration method, the numerical scheme is only approximately satisfied at each solver step, and the degree of violation eventually depends on the weighting constant $\alpha$ in the cost function. The tuning of the weight $\alpha$ is thus more critical  as compared to truncated simulation error minimization.


\section{Case studies}
\label{sec:example}
The performance of the model structures and fitting algorithms introduced in this paper are tested  on  three cases studies considering the identification of a nonlinear RLC circuit; a  \emph{Cascaded Tanks System} (CTS); and  an \emph{Elecro-Mechanical Positioning System} (EMPS). 

\paragraph{Code availability} 
The software implementation is based on the \emph{PyTorch} Deep Learning Framework \citep{paszke:2017automatic}.
All the codes required to reproduce the results reported in the paper are available in the on-line repository \url{https://github.com/forgi86/sysid-neural-continuous}. 

\paragraph{Dataset availability} 
The RLC circuit dataset is simulated and available in the on-line repository. Experimental datasets for the CTS and the EMPS case studies are obtained from the  website \url{http://www.nonlinearbenchmark.org}, which hosts a collection of public benchmarks widely used in system identification. 

\paragraph{Metrics}
 For all case studies,  the performance of the fitting algorithms is assessed in terms of the $R^2$ index of the model simulation:
\begin{equation*}
R^2 = 1 - \frac{\sum_{k=0}^{\nsamp-1} \left(y_k - \simul{y}(t_k)\right)^2}{\sum_{k=0}^{\nsamp-1} \left(y_k - y^{\rm mean}\right)^2},
\end{equation*}
where $y^{\rm mean} = \frac{1}{N}\sum_{k=0}^{\nsamp-1} y_k$.  

For the CTS, the \emph{Root Mean Squared Error} (RMSE) of the model simulation is also provided, as this is the performance index suggested  in the description of the benchmark~\cite{schoukens2017three}:
\begin{equation*}
 \mathrm{RMSE} = \sqrt{\frac{1}{\nsamp}\sum_{k=0}^{\nsamp-1} \left(y_k-\simul{y}(t_k)\right)^2}.
\end{equation*}

The performance indexes are evaluated both on the training dataset and on a separate test dataset. In the case of systems with multiple output channels, the metrics are computed channel-wise.
{
\paragraph{Data pre-processing} For the EMPS example, basic data pre-processing steps described in the corresponding section have been applied on the training dataset to improve the algorithm performance.
In the other cases, the original training datasets have been used for training as the available signals are already in a reasonable numerical range to apply our methods.
Even though pre-processing steps may provide marginal performance improvements also in these 
examples, we decided to present the results obtained using the original datasets for the sake of simplicity.
In all examples, the metrics and plots are reported for variables in the original unit of measure.
}
\paragraph{Algorithm settings} 
In truncated simulation error minimization (Algorithm~\ref{algo:minibatch_multistep_simulation}), the batch size $\batchsize$ and sequence length $\seqlen$    are chosen within the integer range $[32\; 128]$, while the regularization  weight  $\alpha$ in~\eqref{eqn:Jtot} is always set to 1.

In the soft-constrained integration method (Algorithm~\ref{algo:minibatch_consistency_method}), we consider instead a single subsequence 
containing all the dataset samples, \emph{i.e.}, $\batchsize=1$ and $\seqlen=\nsamp$. The weight constant $\alpha$ is  tuned based on the simulation performance of the identified model  in the training  dataset. 

In all the examples, the Adam optimizer is used for gradient-based optimization. The learning rate $\lambda$ is adjusted through a rough trial-and-error within the range $[10^{-2}\; 10^{-6}]$, while the other optimizer parameters are left to default. The number of training steps $\numiter$ is chosen sufficiently high to reach a cost function plateau.

The neural networks' weight parameters are initialized to random Gaussian 
variables with zero mean and standard deviation $10^{-4}$, while the bias terms are initialized to zero. The hidden state variables $\hidden{X}$ are initialized differently for the three examples, exploiting available process knowledge.

\paragraph{Hardware configuration} All computations are carried out on a PC equipped with an Intel i5-7300U 2.60~GHz processor and 32 GB of RAM.

 \subsection{Nonlinear RLC circuit}
\begin{figure}
\centering
\begin{subfigure}{.45\textwidth}
  \centering
  \includegraphics[width=1.0\linewidth]{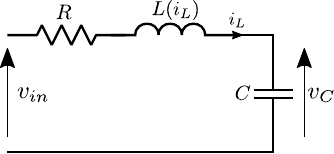}
  \label{fig:sub1}
\end{subfigure}%
\begin{subfigure}{.45\textwidth}
  \centering
  \includegraphics[width=.99\linewidth]{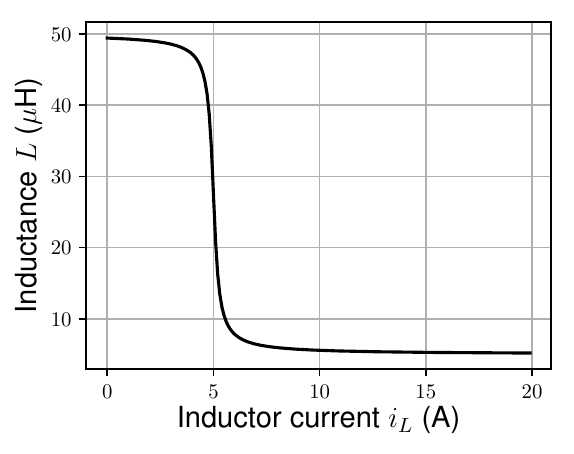}
  \label{fig:sub2}
\end{subfigure}
\caption{Nonlinear series RLC circuit used in the example (left) and nonlinear dependence of the inductance $L$ on the  inductor current $i_L$  (right).}
\label{fig:RLC}
\end{figure}

We consider the nonlinear RLC circuit in Figure~\ref{fig:RLC} (left).
The circuit behavior is described by the continuous-time state-space equation 
\begin{equation}
\label{eq:RLC_sys}
\begin{bmatrix}
\dot v_C\\
\dot i_L
\end{bmatrix} = 
\begin{bmatrix}
  0           & \tfrac{1}{C}\\
 \tfrac{-1}{L(i_L)} & \tfrac{-R}{L(i_L)}\\
\end{bmatrix}
\begin{bmatrix}
v_C\\
i_L
\end{bmatrix} +
\begin{bmatrix}
0\\
\tfrac{1}{L(i_L)}
\end{bmatrix} 
v_{in},
\end{equation}
where $v_{in}~\rm(V)$ is the input voltage; $v_C~\rm(V)$ is the capacitor voltage; and $i_L~\rm(A)$ is the inductor current. 
The circuit parameters $R=3~\Omega$ and $C=270$~nF are fixed, while the inductance $L$ depends on $i_L$ as shown in Figure~ \ref{fig:RLC} (right) and according to
\begin{equation*}
 L(i_L) = L_0\bigg[0.9\bigg(\frac{1}{\pi}\arctan\big(-\!5(|i_L|-5\big) + 0.5\bigg) + 0.1 \bigg], 
\end{equation*}
with $L_0=50~{\mu \rm  H}$. This kind of dependence  is typically encountered in ferrite inductors 
operating in partial saturation \citep{dicapua:2017ferrite}.

In this case study, we assume both state variables $v_C$ and $i_L$ to be measured. 
A training dataset $\Did$ with $N=4000$ samples  is built by  
simulating the system for  $2~\text{ms}$ with a fixed step $\Delta t=0.5~\mu \text{s}$.  The input $v_{in}$ is a filtered white noise with bandwidth $150~\text{kHz}$ and standard deviation $80~\text{V}$. An independent test dataset is generated using as input $v_{in}$ a filtered white noise   with  bandwidth $200~\text{kHz}$ and standard deviation $60~\text{V}$. 
In the training dataset, the observations of $v_C$ and $i_L$ are corrupted by an additive white Gaussian noise with zero mean and standard deviation $10~\rm V$ and $1~\rm A$, respectively. This corresponds to a \emph{Signal-to-Noise Ratio} (SNR) of  20 dB and 13 dB on  $v_C$ and $i_L$, respectively.

\begin{figure}[!t]
\centering
   \includegraphics[width=.75\linewidth]{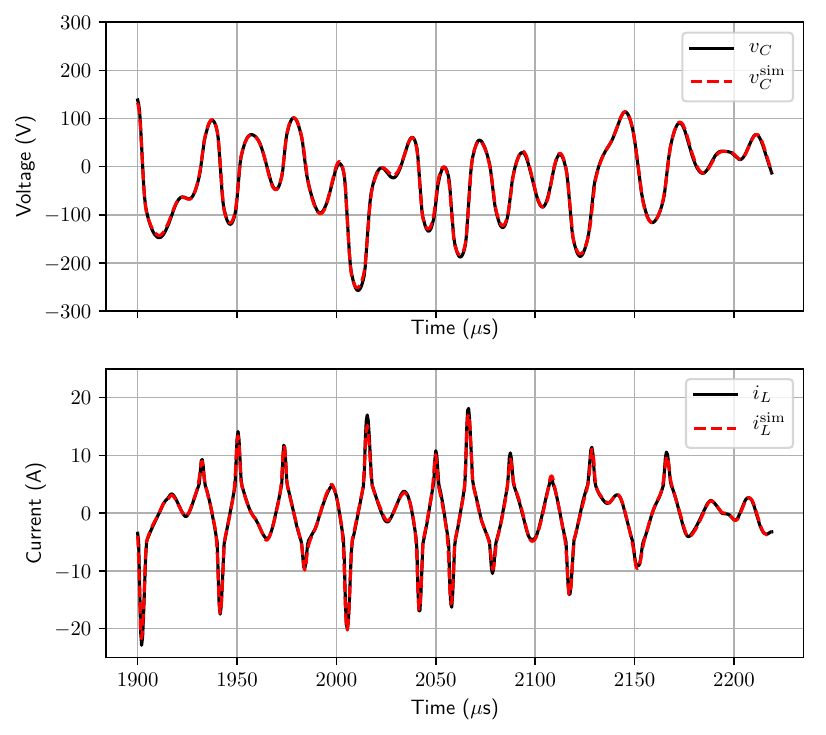}
\caption{RLC circuit: true output (black) and  estimated output (red) obtained by the state-space model trained by truncated simulation error minimization.}
\label{fig:RLC_SS_val_64step_noise}
\end{figure}
\paragraph{Neural Model Structure}
By considering as state vector $x = [v_C\; i_L]^\top$ and input $u=v_{in}$, we adopt for this system the fully observed state model structure \eqref{eq:neural_ss_full} introduced in 
Section \ref{sec:model_structure} reported for convenience below: 
\begin{align*}
\dot x
 & = \NN_{\!f}(x, u) \\
 y &=  
 x.
\end{align*}
This model structure embeds the knowledge that $(i)$ the system has a second-order state-space representation and $(ii)$ the whole state vector is measured. 

The neural network $\NN_{\!f}$ used in this example has feed-forward structure with three input units (corresponding to $v_C$, $i_L$, and $v_{in}$); a hidden layer with 64 linear units followed by ReLU nonlinearity; and two linear output units---the components of the state equation to be learned. 

\paragraph{Truncated simulation error minimization}
Algorithm~\ref{algo:minibatch_multistep_simulation} is executed with learning rate $\lambda=10^{-4}$, number of iterations is $\numiter=10000$, and 
batches containing $\batchsize=64$ subsequences, each one of size $\seqlen=64$. 
The hidden state variables $\hidden{X}$ are initialized to the values of the noisy output measurements. 
Model equations \eqref{eqn:TS:Cauchy} are numerically integrated using the forward Euler numerical scheme. 
The total run time of Algorithm \ref{algo:minibatch_multistep_simulation} is 142 seconds. 

Time trajectories of the true and model output are reported in Figure~\ref{fig:RLC_SS_val_64step_noise}. For the sake of visualization, only a portion of the test dataset is shown.   
The fitted model describes the system dynamics with high accuracy. On both the training and the test datasets, the $R^2$ index in simulation is above $0.99$ for $v_C$ and $0.98$ for $i_L$. For the sake of comparison, a second-order linear \emph{Output Error} model estimated using the \emph{System Identification Toolbox}  \citep{ljung2012version} achieves an $R^2$ index of $0.96$ for $v_C$ and $0.77$ for $i_L$ on the training dataset, and $0.94$ for $v_C$ and $0.76$ for $i_L$ on the test dataset. 

{
Table \ref{tbl:window_sensitivity} reports the run-time and the achieved $R^2$ index in test on $v_C$/$i_L$ obtained by re-running truncated simulation error minimization algorithm with different values of the subsequence length $m$, while keeping all other settings constant. The run-time depends approximately linearly on $m$. This trend is more clear for the larger values of $m$ considered. The model performance generally improves with increasing values of $m$, with the exception of a small performance decrease observed for $m=256$.
}

\begin{table}
{
\begin{tabular}{c|cc}
 Subsequence length $m$ (-) & Run-time (s)&  $R^2$ index (-)\\ 
\cline{1-3}
  2  & 16 & 0.95/0.87 \\
  4  & 19 &  0.98/0.92 \\
  8  & 27&  0.99/0.97  \\
  64  & 138&  0.99/0.98  \\
  128  & 268&  0.99/0.99  \\
  256   &536 & 0.99/0.98
\end{tabular}
\caption{{RLC circuit: run-time and $R^2$ index in test on the voltage/current channels for truncated simulation error minimization with increasing value of the subsequence length $m$.}}
\label{tbl:window_sensitivity}
}
\end{table}


\paragraph{Full simulation error minimization}
Full simulation error minimization is also tested. This method yields the same performance of 64-step truncated simulation error minimization in terms of $R^2$ index of the fitted model. However, the run-time required to execute $\numiter=10000$ iterations and reach a cost function plateau is about two hours.

\paragraph{Soft-constrained integration}
{
The soft-constrained integration method described in Algorithm~\ref{algo:minibatch_consistency_method} is tested.
The regularization constant $\alpha$ and the learning rate $\lambda$ are set to $100$ and $10^{-3}$, respectively.
The algorithm run-time for $\numiter=20000$ gradient descent iterations is 135~seconds.
}
{The estimated model achieves $R^2$ index of $0.99$ and $0.98$ for $v_C$ and $i_L$, both in the training and in the test datasets.
}

\paragraph{One-step prediction error minimization}
For the fully-observed neural model structure, a straightforward fitting criterion may be defined by taking the noisy output measurement as a state estimate and by minimizing the \emph{one-step prediction error loss}
\begin{equation}
\label{eq:J_pred}
 J_{\rm pred}(\theta) = \sum_{t=1}^{\nsamp-1} \norm{y_{t} - y_{t-1} - \Delta t \NN_f({y_{t-1}, u_{t-1}})}^2.
\end{equation}
Minimization of \eqref{eq:J_pred} corresponds to the training of a \emph{standard} feedforward neural network with target $\frac{y_{t} - y_{t-1}}{\Delta t}$ and features $y_{t-1}$, $u_{t-1}$.

\begin{figure}[!t]
\centering
   \includegraphics[width=.75\linewidth]{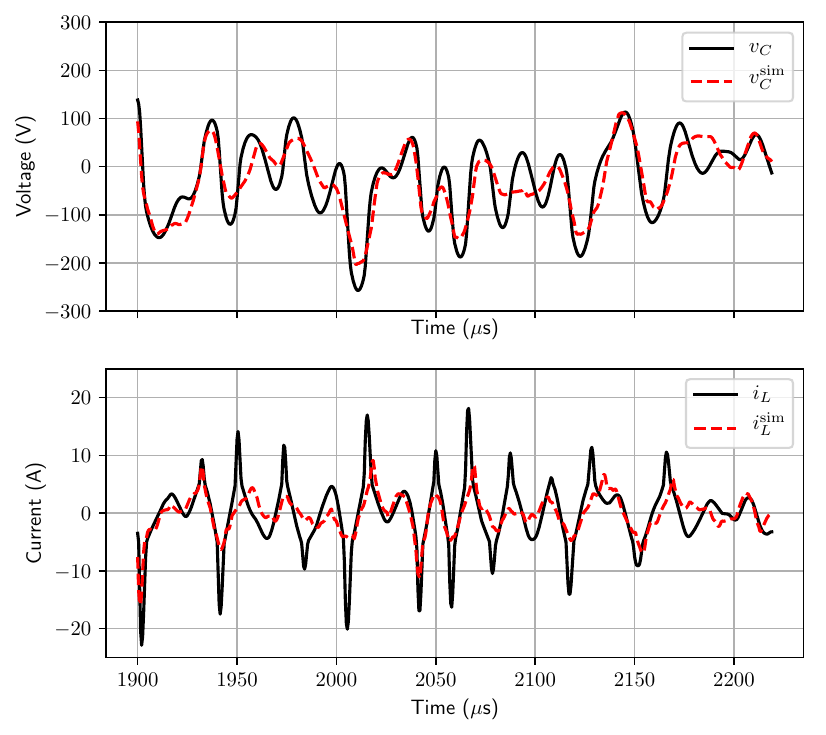}
\caption{RLC circuit: true output (black) and  estimated output (red) obtained by the state-space model trained by one-step prediction error minimization.}
\label{fig:RLC_SS_val_1step_noise}
\end{figure}

Since the neural network is fed with noisy input data and only the one-step ahead is minimized, this approach is not robust to the measurement noise, as illustrated by the authors in \citep{forgione20Uz}.
On this RLC example, the fully-observed state neural model structure trained by minimizing $J_{\rm pred}(\theta)$ achieves an $R^2$ index of $0.73$ for $v_C$ and $0.03$ for $i_L$ in the test dataset (see time trajectories in Figure~\ref{fig:RLC_SS_val_1step_noise}).

By repeating the fitting procedure on a noise-free training dataset, one-step prediction error minimization recovers the same performance of simulation error minimization and soft-constrained integration ($R^2$ index of $0.99$ and $0.98$ for $v_C$ and $i_L$, respectively).

{
\paragraph{Noise sensitivity}
For the proposed truncated simulation error minimization and soft-constrained integration approaches, the result of a sensitivity analysis with respect to  increasing levels of measurement noise is summarized in Table \ref{tbl:noise_sensitivity}. In particular, we consider the cases where the measurement noise on the two channels is multiplied by factors 0x, 1x, 2x, 3x, and 6x. Note that the factors 0x and 1x correspond to the noise-free and to the noisy datasets considered above.}

{
The truncated simulation error minimization approach is clearly less sensitive to measurement noise w.r.t. to soft-constrained integration. Still, soft-constrained integration is more noise-tolerant than the one-step prediction error approach, which is only applicable to the noise-free dataset (0x), as shown in the previous paragraph.
}

\begin{table}
{
\begin{tabular}{c|cccccc}
\multicolumn{1}{}{} & \multicolumn{5}{c}{Noise level} \\
 Algorithm  & 0x    &  1x   & 2x    & 3x    & 6x   \\ 
\cline{1-6}
  TSEM  & 0.99/0.98 &  0.99/0.98 &  0.99/0.97 &  0.98/0.94 &  0.96/0.91\\
  SCI  & 0.99/0.99 &  0.99/0.98 &  0.89/0.7 &  0.38/-0.027 &  0.035/-0.23 &  \\
\end{tabular}
}
\caption{
{RLC circuit: $R^2$ index for the voltage/current channels on the test set for truncated simulation error minimization (TSEC, first row) and soft-constrained integration (SCI, second row) for increasing noise level.}}
\label{tbl:noise_sensitivity}
\end{table}

\subsection{Cascaded Tanks System} 
We consider the CTS already introduced in Section \ref{sec:model_structure} and
 described  in details in \citep{schoukens2017three}.  

The training and test datasets contain 1024 points each, collected at a constant sampling time $\Delta t=5~\rm{s}$. In both datasets, the input is a multisine signal with identical power spectrum, but different realization. 
Input and output values are in Volts as they correspond to the actuator commands and the raw sensor readings, respectively.

The initial state of the system is not provided, but it is known to be the same for both datasets. Thus, as suggested in \citep{schoukens2017three}, 
we use the initial state estimated on the training dataset (which is a by-product of the proposed fitting procedures) as initial state for model simulation in test.

\paragraph{Neural Model Structure}
The neural model structure used for this system is 
\begin{subequations}
\label{eq:neural_CTS_overflow}
\begin{align}
 \dot x_1 &= \NN_{\!f_1}(x_1, u) \\
 \dot x_2 &= \NN_{\!f_2}(x_1, x_2, u) \\
      y   &= x_2.
\end{align}
\end{subequations}
Compared to \eqref{eq:neural_CTS}, model \eqref{eq:neural_CTS_overflow} also includes a direct dependency on $u$ in the second state equation. This dependency is added to take into account that in this experimental setup, in case of water overflow from the upper tank, part of the overflowing water may go directly in the lower tank \citep{schoukens2017three}.

The neural networks $\NN_{\!f_1}$ and $\NN_{\!f_2}$ have two and three input units, respectively. Both networks one hidden layer with 100 linear units followed by ReLU nonlinearity and a linear output unit.

\paragraph{Truncated Simulation Error Minimization}
Algorithm~\ref{algo:minibatch_multistep_simulation} is executed with learning rate $\lambda=10^{-3}$, number of iterations $\numiter=10000$,  batch size $\batchsize=64$, and subsequence length $\seqlen=128$. 
The  components of the hidden state variables $\hidden{X}$ associated to the state variable   $x_1$ are initialized to $0$, while the  components associated  to $x_2$ are initialized to the noisy output measurements $y$. Model equations~\eqref{eq:cauchy_approx} are numerically integrated using the forward Euler numerical scheme. 
The total run-time of Algorithm~\ref{algo:minibatch_multistep_simulation} is 533 seconds.

\begin{figure}
\centering
\begin{subfigure}{.5\textwidth}
  \centering
  \includegraphics[width=\linewidth]{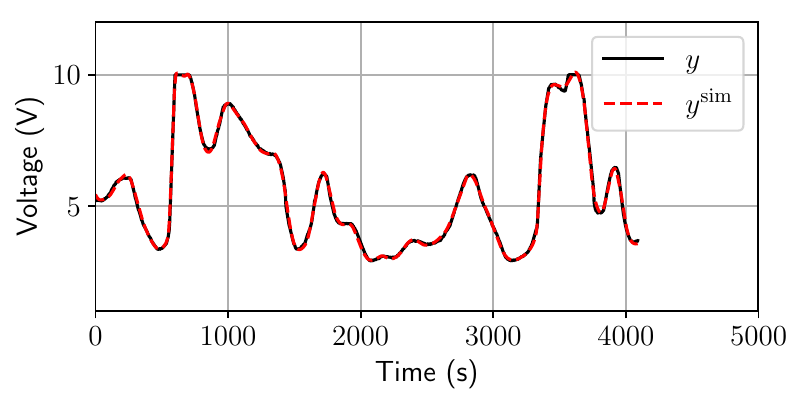}
  \caption{Training dataset.}
  \label{fig:sub1}
\end{subfigure}%
\begin{subfigure}{.5\textwidth}
  \centering
  \includegraphics[width=\linewidth]{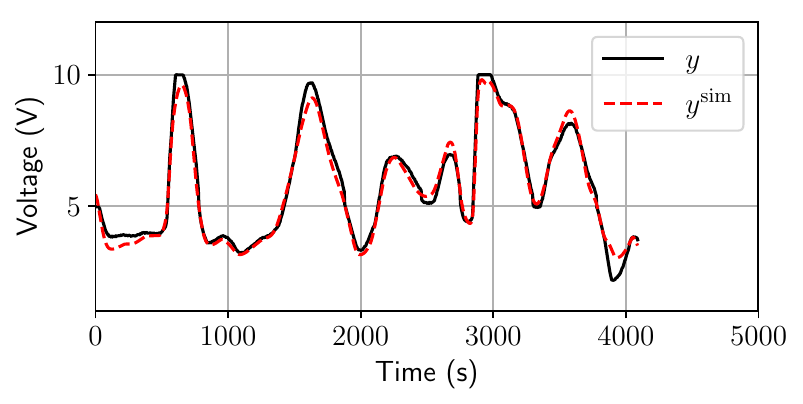}
  \caption{Test dataset.}
  \label{fig:sub2}
\end{subfigure}
\caption{CTS: measured output $y$ (black) and model simulation $y^{\rm sim}$ (red) obtained by the neural model trained by truncated simulation error minimization.} 
\label{fig:CTS_model_custom_SS_128step.pdf}
\end{figure}

Time trajectories of simulated and true output  are reported in Figure~\ref{fig:CTS_model_custom_SS_128step.pdf}. 
The achieved $R^2$ index  is $0.99$ and $0.97$ on the training and on the test dataset, respectively. The RMSE index is $0.08$~V and $0.33$~V on the identification and on the test dataset, respectively.
These results compare favorably with state-of-the-art black-box nonlinear identification methods applied to this benchmark \citep{relan2017unstructured, birpoutsoukis2018efficient, svensson2017flexible}. 
{To the best of our knowledge, the best previously published result was obtained in \cite{relan2017unstructured} using non-linear 
basis expansion state-space models and exploiting a state initialization procedure (called NLSS2) based on a preliminary (linear) subspace identification for optimization. The authors of \citep{relan2017unstructured} report an RMSE index of $0.34$~V on the test dataset.}
A higher performance  was reported  only in \citep{rogers2017grey} for grey-box models including an explicit physical description of the water overflow phenomenon (RMSE index of $0.18$~V). 
 
Note that the largest discrepancies between measured and simulated output are noticeable towards the end of the test experiment, where the measured output $y$ is close to $2$~V. This condition is not encountered in the training dataset and therefore a mismatch can be expected for a  black-box nonlinear  model such as a neural network.
	
\paragraph{Soft-constrained integration method}
Algorithm~\ref{algo:minibatch_consistency_method} is executed with the regularization term $J_{\rm reg}$ in~\eqref{eq:soft_crank-nicolson} enforcing the Crank-Nicolson integration scheme  and a regularization constant $\alpha=50000$.
For this small dataset, all time steps fit into the memory and can be processed altogether in a batch. Thus, we consider batches with  a single subsequence ($\batchsize=1$) containing the whole training dataset $\Did$ $(\seqlen = 1024)$. Optimization is performed over $\numiter=50000$ iterations of the Adam algorithm, with learning rate $\lambda=10^{-5}$. The total run-time of Algorithm~\ref{algo:minibatch_consistency_method} is 271 seconds---approximately half the run-time of Algorithm~\ref{algo:minibatch_multistep_simulation}.

\begin{figure}
\centering
\begin{subfigure}{.5\textwidth}
  \centering
  \includegraphics[width=\linewidth]{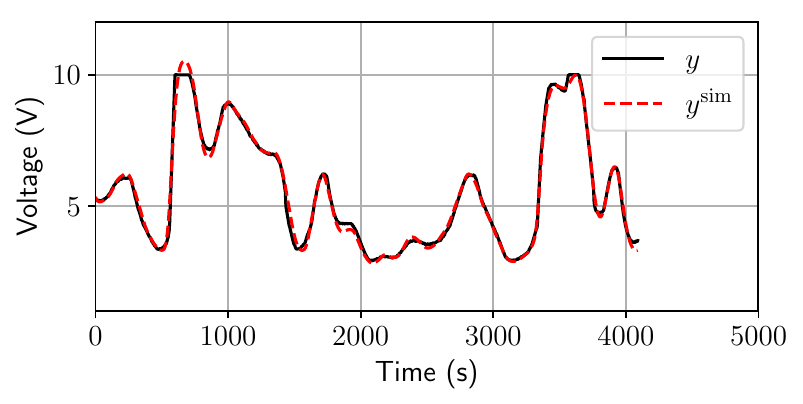}
  \caption{Training dataset.}
  \label{fig:sub1}
\end{subfigure}%
\begin{subfigure}{.5\textwidth}
  \centering
  \includegraphics[width=\linewidth]{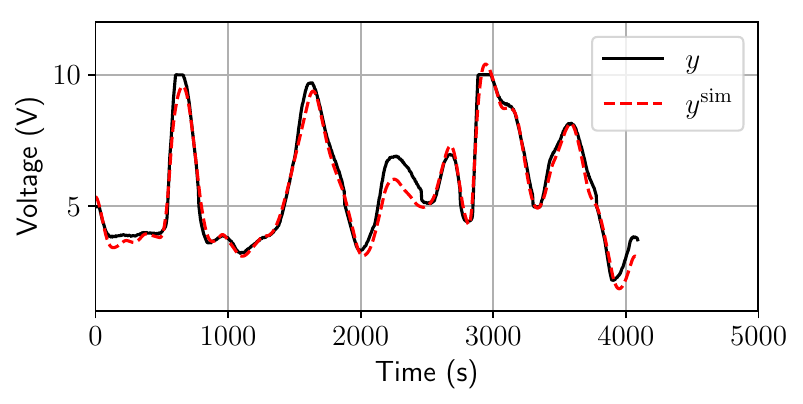}
  \caption{Test dataset.}
  \label{fig:sub2}
\end{subfigure}
\caption{CTS: measured output $y$ (black) and model simulation $y^{\rm sim}$ (red) obtained by the neural model trained using the soft-constrained integration method.} 
\label{fig:CTS_model_SS_custom_consistency.pdf}
\end{figure}

Time trajectories of the output  are reported in Figure~\ref{fig:CTS_model_SS_custom_consistency.pdf} for both the training and the test dataset. 
The $R^2$ index of the model is $0.99$ and $0.96$ on the training and on test dataset, respectively. The RMSE index is $0.18$~V and $0.40$~V on the training and on the test datasets, respectively. The results are thus in line with the ones achieved by truncated simulation error minimization. 

{Table \ref{tab:CTS_perf_comp} summarizes the results obtained in the literature on the
CTS benchmark, together with the ones of truncated simulation error minimization and soft-constrained integration in terms of the $\rm{RMSE}$ index in test.}

	\begin{table}
		\centering
		{
		\begin{tabular}{ l | c   }
			Method & $\rm{RMSE}$ (V) \\
			\hline
			Best linear approximation \cite{relan2017unstructured} & 0.75\\
			Volterra model \cite{birpoutsoukis2018efficient} & 0.54\\
			State-space with GP-inspired prior \cite{svensson2017flexible} & 0.45\\
			Non-linear SS + NLSS2 initialization \cite{relan2017unstructured} & 0.34\\
			Grey-box with physical overflow model \cite{rogers2017grey} & 0.18\\
			\hline
			TSEM [this paper]  &  0.33\\
			SCI [this paper]  &  0.40\\
		\end{tabular}
		}
		\caption{{CTS: comparison of truncated simulation error minimization (TSEM) and soft-constrained integation (SCI) with
		other published methods in the literature in terms of the $\rm{RMSE}$ index in the test dataset.}}
		\label{tab:CTS_perf_comp}
	\end{table}

\subsection{Electro-Mechanical Positioning System}
As a last case study, we consider the identification of the Electro-Mechanical Positioning System (EMPS)  described in \citep{janot2019data}. 

The system is a controlled prismatic joint, which is a common component of robots and machine tools. 
In the benchmark, the system input is the motor force $\tau$~(N) expressed in the load side and the measured output is the prismatic joint position $p$~(m).
A physical state-space model for the system is
\begin{subequations}
\label{eq:physical_EMPS}
\begin{align}
 \dot p &= v \label{eq:physical_EMPS_pos}\\
 \dot v &= -\frac{\tau}{M} - \frac{f_v}{M} v  - \frac{F_c(v)}{M}, \label{eq:physical_EMPS_velo}
\end{align}
\end{subequations}
where $M$~(kg) is the joint mass, $f_v$~(Ns/m) is the dynamic friction coefficient and $F_c$~(N) is the static friction.

The identification and test dataset are constructed from closed-loop experiments performed with the same reference position trajectory. A force disturbance is acting on the system in the test experiment only.
The two datasets have the same duration (approximately 25 seconds) and are collected at a sampling frequency of $1$~kHz.  

{For training, the original EMPS signals  are decimated by a factor $5$. Thus, each dataset contains $\nsamp=4968$ points with sampling time $\Delta t=5$~ms.} {Furthermore, the output position is normalized to the numerical range $[-1\ 1]$. Finally, an initial estimate of the joint velocity to be used for initialization of the hidden state variables is obtained by numerical differentiation (using a forward finite differences scheme) of the output position signal.}

\paragraph{Neural Model Structure}
According to the physical model~\eqref{eq:physical_EMPS}, the neural model structure used to fit the EMPS system is
\begin{subequations}
\label{eq:neural_EMPS}
\begin{align}
 \dot x_1 &= x_2 \\
 \dot x_2 &= \NN_{\!f}(x_2, u) \\
      y   &= x_1,
\end{align}
\end{subequations}
  with state variables   $x_1 = p$ and $x_2=v$; and  input $u=\tau$.  
  
The neural model structure~\eqref{eq:neural_EMPS}  
captures the physical knowledge that: $(i)$ the system states are position and velocity; $(ii)$ 
the derivative of position is velocity; and $(iii)$ the velocity dynamics does not depend on the position. The neural network is thus used to describe the velocity dynamics \eqref{eq:physical_EMPS_velo}, which could be rather complex due to the presence of static friction. Indeed, static friction is highly nonlinear and hard to describe with first-principles formulas. On the other hand, there is no need to use a black-box model to describe the position dynamics \eqref{eq:physical_EMPS_pos}. In fact, this equation simply states that velocity is the time derivative of position.

The neural network $\NN_{\!f}$ used for this benchmark has 2 input units; 64 hidden linear units followed by ReLU nonlinearity; and one linear output unit.
\paragraph{Truncated simulation error method}
Algorithm~\ref{algo:minibatch_multistep_simulation} is executed with learning rate $\lambda=10^{-4}$, number of iterations is $\numiter=10000$,   batch size $\batchsize=32$, and sequence length $\seqlen=64$. 
{The components of $\hidden{X}$ associated to $x_1$ are initialized to the position sequence, while the components associated to $x_2$ are initialized using the initial velocity estimate obtained through numerical differentiation of the position sequence in the pre-processing step.}

Model equations~\eqref{eq:cauchy_approx} are numerically integrated using  the  fourth-order Runge-Kutta scheme RK44 \citep{ralston1962runge}. The total run-time of Algorithm~\ref{algo:minibatch_multistep_simulation} is $710$~seconds.

\begin{figure}
 \begin{subfigure}{.5\textwidth}
   \centering
   \includegraphics[width=\linewidth]{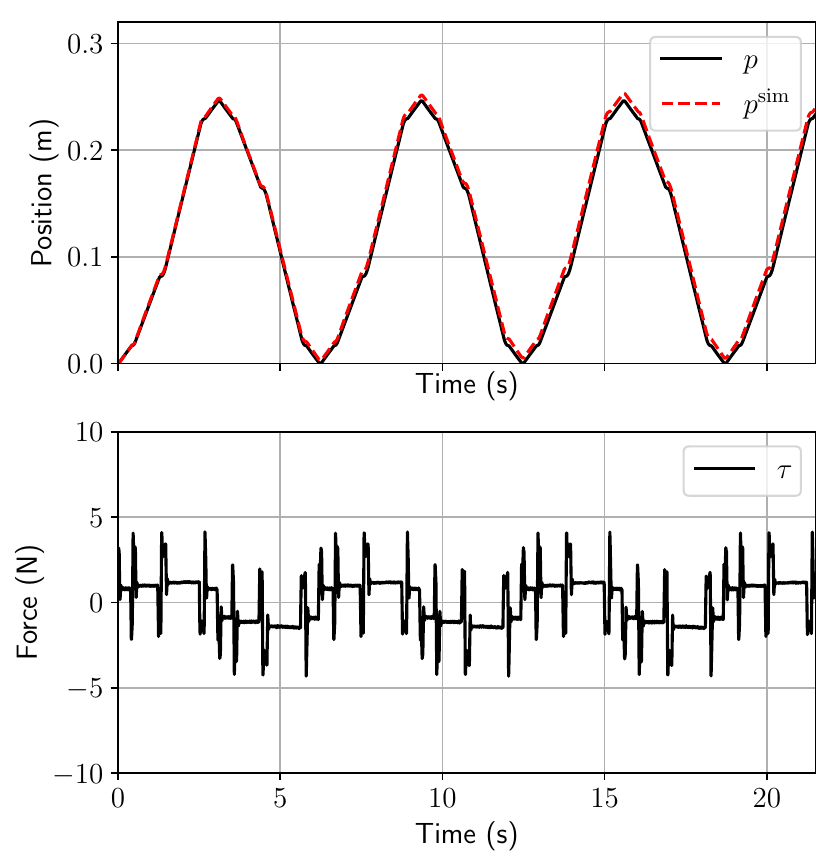}
   \caption{Training dataset.}
   \label{fig:sub1}
 \end{subfigure}%
 \begin{subfigure}{.5\textwidth}
   \centering
   \includegraphics[width=\linewidth]{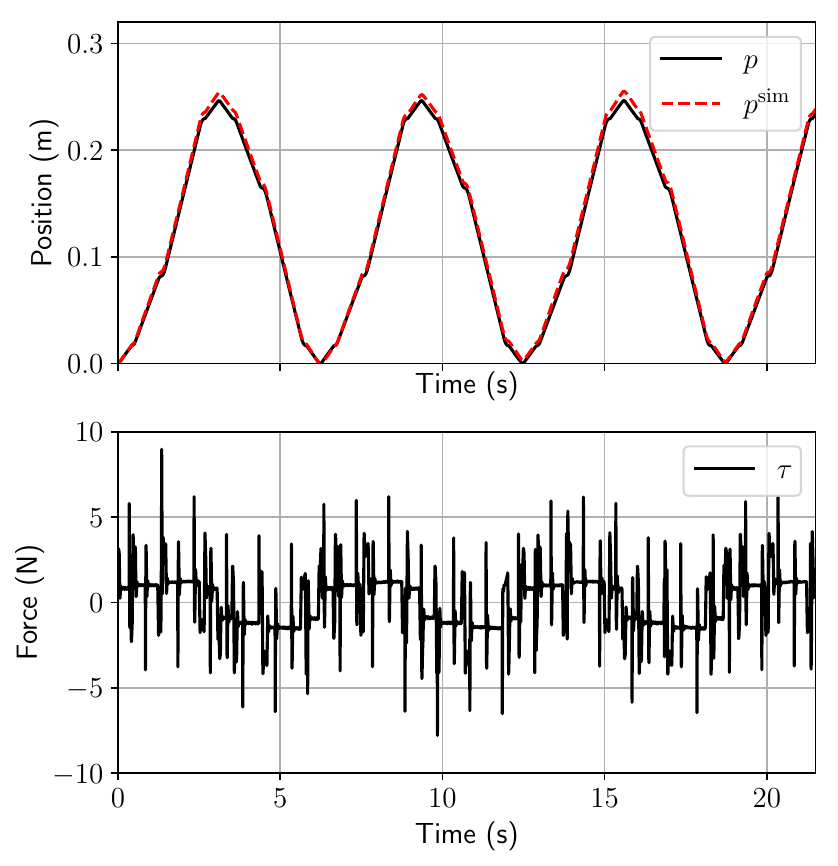}
   \caption{Test dataset.}
   \label{fig:sub2}
 \end{subfigure}
  \caption{EMPS: measured position $p$ (top panels, black) and model simulation $p^{\rm sim}$ (top panels, red) obtained by the neural model trained by truncated simulation error minimization. The input force $\tau$ is shown in the bottom panels.} \label{fig:EMPS}
 \end{figure}

Time trajectories of the input and of the output  are reported in Figure~\ref{fig:EMPS}.
The achieved $R^2$ index is above $0.99$ on both the identification and test datasets. By comparison, \cite{janot2019data} reports an $R^2$ index of $0.5$ for different linear models
estimated on this benchmark. 

\paragraph{Soft-constrained integration method}
Algorithm~\ref{algo:minibatch_consistency_method} is executed with the regularization term $J_{\rm reg}$ in~\eqref{eq:soft_backward_euler} enforcing the backward Euler integration scheme and regularization weight $\alpha=1000$. As for the CTS benchmark, we consider batches with a single subsequence ($\batchsize=1$) containing the whole training dataset $\Did$ $(\seqlen = 4968)$. 
Optimization is performed over $\numiter=40000$ iterations of the Adam algorithm, with learning rate $\lambda=10^{-5}$. 
The identified model achieves an $R^2$ index above $0.99$ both in identification and in test, as for Algorithm~ \ref{algo:minibatch_multistep_simulation}.
 The total run-time of Algorithm~\ref{algo:minibatch_consistency_method} is $ 342$~seconds (around 2x faster then Algorithm~\ref{algo:minibatch_multistep_simulation}).

\section{Conclusions and follow-up}
\label{sec:conclusions}
%
%
%

In this paper, we have presented neural model structures and two novel methodologies for the identification of continuous-time dynamical systems.

The main strengths of the presented framework are: $(i)$ its versatility to describe complex and structured non-linear systems, thanks to the neural network flexibility and the possibility to exploit physical model structures; $(ii)$ its robustness to the measurement noise, thanks to the minimization of a (truncated) simulation error criterion and  regularization terms  that enforce the hidden state variables to be consistent  with the estimated neural model; $(iii)$  the possibility to exploit parallel computing to train the network and optimize the initial conditions, thanks to the division of the dataset into small-size subsequences.  

The proposed case studies have shown the effectiveness of the presented methodologies, and their capabilities to outperform state-of-art algorithms in well-known benchmarks for system identification.
 
Current and future research activities are devoted to the application of the proposed framework  for the identification of  systems described by partial differential equations, as well as the formulation  of alternative fitting criteria that directly take into account the final usage the estimated dynamical models, like fault detection and control system design.

\section*{Acknowledgements}
This work was partially supported by the European H2020-CS2 project ADMITTED, Grant agreement no. GA832003.

\bibliography{ms}

\end{document}